# Flexible antiferromagnetic FeRh tapes as memory elements

Ignasi Fina,[†,*] Nico Dix,[†] Enric Menéndez,[‡] Anna Crespi,[†] Michael Foerster,[§] Lucia Aballe,[§] Florencio Sánchez,[†] Josep Fontcuberta[†,*]

[†]Institut de Ciència de Materials de Barcelona (ICMAB-CSIC), Campus UAB, E-08193 Bellaterra, Catalonia, Spain.

[‡]Departament de Física, Universitat Autònoma de Barcelona, E-08193 Bellaterra, Catalonia, Spain.

[§]ALBA Synchrotron Light Facility, Carrer de la Llum 2-26, Cerdanyola del Vallès, Barcelona 08290, Spain.

ABSTRACT: The antiferromagnetic to ferromagnetic transition occurring above room temperature in FeRh is attracting interest for applications in spintronics, with perspectives for robust and untraceable data storage. Here, we show that FeRh films can be grown on a flexible metallic substrate (tape shaped), coated with a textured rock-salt MgO layer, suitable for large scale applications. The FeRh tape displays a sharp antiferromagnetic to ferromagnetic transition at about 90 ºC. Its magnetic properties are preserved by bending (radii of 300 mm), and their anisotropic magnetoresistance (up to 0.05 %) is used to illustrate data writing/reading capability.

**1. Introduction**

Development of functional material on flexible substrates is a fast growing activity due to enormous potential for applications as technology moves towards mobile and wearable. Electronic components on flexible substrates have already been receiving attention in many other high-tech areas,[1-2] including sensing,[3] biomedical,[4] or photovoltaics.[5-6] Referring to magnetic sensing, giant magnetoresistance and magnetoimpedance devices are probably ahead in the race.[7-9] Most magnetic materials and devices have been engineered on polymeric substrates. For instance, embedded ferromagnetic elements,[10-11] anisotropic magnetoresistive,[12] GMR[13] and Hall effect sensors,[14] synthetic antiferromagnets (SAF),[9] magnetic tunneling devices[15], or magnetoelectric devices[16-19] have been reported. In most of the cases, non-ordered magnetic metallic alloys were used because they allow close-to-room-temperature deposition as required by the polymeric substrates used. In contrast, the design of devices exploiting the rich variety of properties of oxides has been limited by the relatively high temperature needed to grow crystalline perovskite or spinel oxides and the scarcity and cost of suitable substrates. Recently, flexible muscovite silicate has been used as template for the growth of crystalline ferromagnetic and ferroelectric oxides[20-25] and other complex compounds,[26] thus opening new perspectives.

Other than sensing, magnetic materials showing ferromagnetic order are of relevance as data storage media and used in hard disks and magnetic random access memories. Reverse of the magnetization direction of a memory cell is usually realized by magnetic fields. By the same token, a spurious magnetic field could erase the stored information. In addition, the stray magnetic fields may induce crosstalk with neighboring magnetic memory cells limiting data storage density. The use of magnetic materials for data storage and computing, so-called spintronics, is now turning the focus towards antiferromagnetic materials. Antiferromagnetic materials show intrinsic zero magnetization, but, being magnetically ordered, they can store information. Indeed, perpendicularly aligned antiferromagnetic elements correspond to different memory states. These two states can be read by electric transport measurements,[27-28] but are indistinguishable if common magnetic reading-heads elements would be used because its null magnetization. Therefore, data can be cloaked. Moreover, the robustness of the antiferromagnetic order implies insensitiveness to spurious magnetic fields and its null magnetization implies absence of crosstalk, subsequently allowing higher data density.[29-30] Applications that would require magnetic robustness and information invisibility such as flexible credit cards could benefit from flexible antiferromagnetic memories. With this aim, it is necessary to find routes that allow mass production of flexible antiferromagnetic materials able to store information.

α-FeRh (in the following, FeRh) is a magnetic alloy antiferromagnetically ordered (AFM) at room temperature. Upon heating to about 50-150°C, this alloy undergoes a phase transition towards a ferromagnetic (FM) state. The uncommon AFM to FM transition, occurring near room temperature, makes FeRh an interesting material for applications. For instance, it has been shown that, after a suitable data writing step in the high temperature FM state, FeRh films can be used to store magnetic information in its AFM state.[31] Thus FeRh is an AFM material able to store information and thus its production as a flexible tape is interesting for applications. However, the growth of FeRh thin films with sharp and full AFM to FM transitions is extremely challenging, because the required full order of Fe/Rh is highly sensitive to different growth conditions, used substrate conditions and epitaxial strain.[32-40] As recently reviewed by Fina et al.,[41] FeRh films have been grown on perovskite, spinel and sapphire substrates.[32-33, 42-43] Optimal quality thin FeRh films have been obtained by sputtering on MgO single crystalline substrates due to the identical rock-salt atomic packaging and reduced structural mismatch (-0.5%).[44-45]

Therefore, the growth of FeRh films on available flexible substrates is a challenging objective. As mentioned MgO is the most suitable substrate. We will take advantage of the availability of MgO-coated flexible conductors made by the so-called inclined substrate deposition technique (ISD), currently used in high-temperature superconducting tape technologies.[46] We will show that, under appropriate growth conditions, FeRh films can be grown on flexible MgO-coated tapes. Results show that the magnetic properties of the films grown on this substrate, rival those of FeRh films grown on single crystalline substrates, displaying a sharp AFM to FM phase transition at about 90 °C. Moreover, we show that films can be reversibly bent at least to strains of about 0.02% without bend-induced cracking. Anisotropic magnetoresistance of our flexible FeRh tapes is used to demonstrate a data storage capability comparable to that reported for FeRh on single crystalline substrates.[31] We end by discussing possible avenues for further development.

## 2. Results

FeRh (57 nm) films were sputtered on a textured MgO coated (≈ 2 μm thick) flexible tape (C267 HASTELLOY®) using the conditions described in the experimental section. The FeRh layers were in-situ capped with 20 nm-Pt to avoid oxidation. The structure of the sample is sketched in **Figure 1**a. To be used as a control sample, a FeRh film was also grown on a single crystalline MgO(001) substrate using identical conditions. In the following, these samples are denoted FeRh-Tape and FeRh-SC, respectively. Other samples were grown with slightly different conditions (Supporting Information Table S1). When appropriate, these complementary samples will used for comparison.

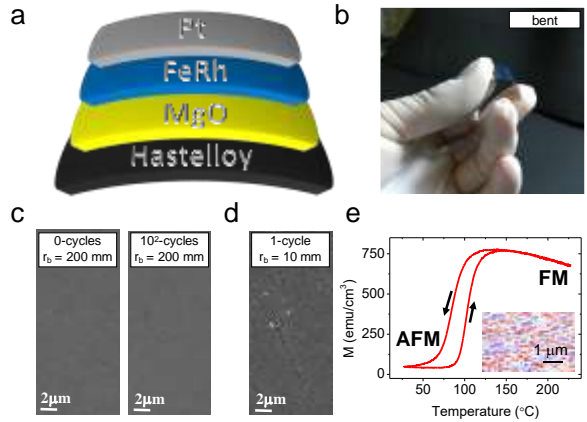

**Figure 1.** (a) Sketch of the flexible FeRh-Tape structure. FeRh film is grown in a Hastelloy tape previously buffered with a MgO textured layer. (b) Photograph of the bent (25 x 12) mm$^2$ FeRh-Tape sample. (c) SE images of representative FeRh-Tape region without being bent after growth (0-cycles) and after 10$^2$ bending cycles up to $r_b$ = 200 mm. (d) SEM image of representative FeRh-Tape region after bending it up to $r_b$ = 10 mm. (e) Magnetization M(T) cycle of the of the FeRh-Tape sample collected at 500 Oe (in-plane applied magnetic field) recorded from RT to 225 ºC (warming) and to RT (cooling) following the arrows directions using VSM. Inset: Ferromagnetic domain distribution at high temperature (T = 150 ºC and H = 0) collected by XMCD-PEEM.

In Figure 1b, we show the picture of a FeRh-Tape 25 mm long, 12 mm wide sample in bent state to illustrate its flexibility. In Figure 1c, we show secondary electrons (SE) scanning electron microscopy (SEM) images of an as-grown FeRh-Tape (left) and after 10$^2$ bending cycles (concave and convex). We used a bending radius of $r_b$ = 200 mm. Intermediate states are shown in Supporting Information Figure S1. No significant damage is observed in the SEM images. In contrast, when the bending radius is reduced to $r_b \approx$ 10 mm, a pattern of cracks, spaced by about 2-5 μm, develops (Figure 1d).

Figure 1e shows the temperature-dependent in-plane magnetization M(T) of the FeRh-Tape sample. In the heating cycle, the M(T) curve displays a sharp increase of magnetization at $T_{N,h} \approx 110$ °C evidencing an AFM to FM phase transition. Upon cooling, the reduction of magnetization (FM to AFM transition) is observed starting at $T_{N,c} \approx 90$ °C. The hysteretic behavior of M(T) is a fingerprint of the first order nature of the AFM to FM phase transition of FeRh. Remarkably, the magnetization value at high temperature and the sharpness of the transition, both being signatures of the film quality, are similar to those observed in the reference FeRh-SC sample (see Supporting Information Figure S2). In the inset of Figure 1e, we include the image of the ferromagnetic in-plane domains (blue and red indicating opposite magnetization direction) collected using X-Ray Magnetic Circular Dichroism in combination with Photoemission Electron Microscopy (XMCD–PEEM) at high temperature (150 ºC). As observed, the size of the ferromagnetic domains (about 1-2 µm) is similar to those of high-quality FeRh films on single crystalline substrates.[47]

In the M(T) data of Figure 1e, it can also be appreciated that at room temperature, in the AFM state, there is a residual magnetization (around 40 emu/cm$^3$), indicating the persistence of a residual FM phase. As expected,[48-50] the residual magnetization is significantly smaller in the FeRh-SC film ($\approx$ 14 emu/cm$^3$). Indeed, the FeRh film grown on the MgO-buffered tape[51] is found to be more more defective than that of the FeRh-SC film, as we shall confirm below. In spite of these probably unavoidable differences, the magnetic data collected on the FeRh-Tape clearly indicate that the α-FeRh phase has been stabilized on the flexible substrate.

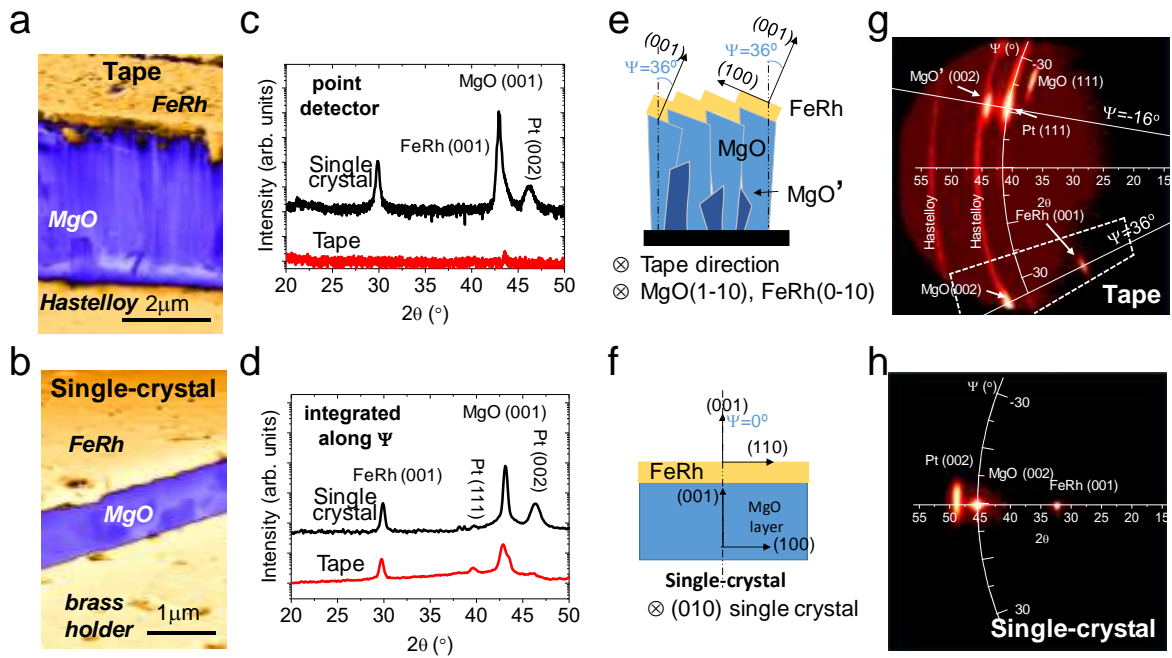

**Figure 2.** (a, b) Tilted SE+BE SEM image of FeRh-Tape and FeRh-SC samples, respectively. (c) θ–2θ scan at the vicinity of the MgO(001) reflection performed with point detector (Normal to the surface (ψ = 0). The diffraction peaks of 00L MgO, Pt and FeRh in the FeRh-SC sample are indexed; all of them are absent in FeRh-Tape. (d) Integrated along ψ intensity versus θ–2θ. (e, f) Sketch of the different crystalline orientation of MgO-Tape and MgO-SC substrates, respectively. (g,h) 2θ–ψ X-ray images along of FeRh-Tape and FeRh-SC, respectively. the FeRh (001) texture of both FeRh layers.

Secondary electron (SE) and Secondary Backscattered (BE) scanning electron microscopy (SEM) images were collected at roughly grazing incidence. **Figure 2**a and Figure 2b are the combined SE and BE images (SE+BE) of the FeRh-Tape and FeRh-SC samples, respectively. In Figure 2a, it can be appreciated that the FeRh-layer of the FeRh-Tape sample has grown with a grainy surface morphology (yellow) on top of a columnar-like MgO template (blue). The biaxial texture of MgO by ISD is well documented.[52-53] Conversely, the surface of the FeRh-SC sample (Figure 2b) is a smooth and continuous FeRh film (yellow) on top of the MgO single-crystal (blue). These obvious microstructural differences account for the dissimilarity of the M(T) data of FeRh-Tape and FeRh-SC samples described above.

Structural analysis was performed using X-Ray diffraction (XRD). We first collected the specular θ-2θ scans shown in Figure 2c using a point detector. The XRD pattern of the FeRh-SC sample displays a strong FeRh (001) reflection in addition to MgO(001) and Pt(002) reflections of the MgO substrate and the Pt protecting layer, respectively. In contrast, no reflections are visible in the specular θ-2θ scans of the FeRh-Tape sample (see Supporting Information Figure S3 for a wider scan). Further analysis (see Supporting Information Figures S4) indicates that FeRh-SC sample shows an epitaxial relationship of [100]FeRh(001)//[110]MgO(001) and the Pt capping layer shows a [100]Pt(001)//[110]FeRh(001) epitaxial relationship.

The absence of MgO(001), Pt(001) and FeRh(001) reflections in symmetric scans of FeRh-Tape sample in Figure 2c, using point detector, is due to its microstructure. ISD produces a columnar biaxially-textured MgO layer with the c-direction tilted away from the surface normal ($\Psi \approx 36º$), which promotes the formation of a roof-tiled surface of nearly atomically flat (001)MgO planes.[54] Thus the (00*l*) reflections of MgO (sketch in Figure 2e)[55] are not accessible in symmetric θ–2θ scans using point detector. Instead, in the FeRh-SC, the MgO(001) planes are normal to the sample surface (Figure 2f) and are visible in the corresponding specular θ-2θ scans (Figure 2c).

To determine the texture of the FeRh-Tape sample, we collected 2θ–ψ frames using a 2D detector (see Supporting Information Figures S5 for experimental set-up configuration and Figures S6 for further data analysis). In Figure 2g, it can be observed that a MgO (001) reflection appears at $\Psi \approx 36º$. Along the same $\Psi \approx 36º$, an intense FeRh(001) reflection is observed. This indicates an epitaxial relation [100]FeRh(001)//[110]MgO(001) as found in the FeRh-SC with the [001] axis of both FeRh and MgO tilted ($\Psi \approx 36°$) as described above. At $\Psi \approx -16°$ of the MgO(002) reflection, the presence of minor fraction of differently oriented MgO grains is visible [labeled MgO'(002)]. It is known that these crystallites are formed at early stage growth of MgO.[56] Near MgO'(002), the MgO(111) reflection corresponding to the main MgO family of crystallites appears at $\Psi \approx 55°$ with respect to the MgO(001), as expected for the (111) and (001) planes in a cubic structure. The reflections of the polycrystalline HASTELLOY are also observed and labelled. It follows that the Pt and FeRh layers grow with (00l) texture on the preferential MgO(001) surface, and maintain the same epitaxial relationships in the FeRh-Tape as in FeRh-SC.

Integrated along $\Psi$ $\theta-2\theta$ scans for FeRh- and FeRh-Tape are similar (Figure 2d) indicating the good crystalline quality of FeRh-Tape sample. Symmetric $2\theta-\Psi$ scans have been also collected on FeRh-Tape samples grown at different conditions (see experimental section and Supporting Information Figure S7) showing similar structural properties. Instead, their magnetic properties differ substantially (Supporting Information Figure S8), reflecting the extreme sensitivity of the FM/AFM balance to the fine details of the FeRh microstructure, not perceptible in XRD data.

We now turn our attention to the effect bending on the magnetic properties of FeRh-Tape. The tape has been laterally pressed by two aluminum pieces to obtain bending radii of $r_b \approx 300$ mm, 500 mm, 1200 mm and $\infty$ ($\equiv$ unbent) as shown in **Figure 3**a.

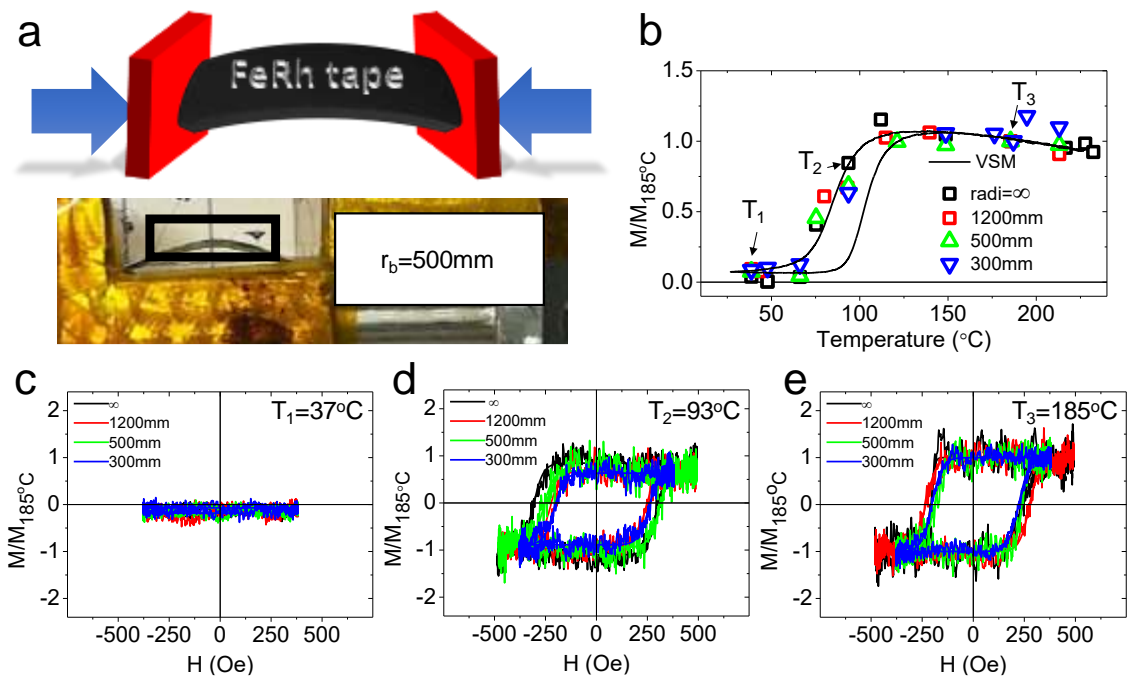

**Figure 3.** (a) Sketch of the experimental set-up used to bend the FeRh-tape together with the lateral pictures of the FeRh-tape bent up to $r_b \approx 500$ mm. The section of the FeRh-Tape sample located within the black rectangle has been used to calculate the $r_b$. (b) MOKE signal normalized to its value at 185 °C measured at several temperatures on FeRh-Tape with different $r_b$. Solid line corresponds to the data collected by VSM on the unbent FeRh sample. (c,d,e) MOKE magnetization versus magnetic field loops recorded at 37, 93 and 185°C, respectively, of bent films.

The variation of the magnetic properties upon bending has been monitored by longitudinal magneto-optic Kerr effect (MOKE) measurements (see the experimental set-up in Supporting Information Figure S9). Data have been collected either at variable temperature upon cooling or isothermally after inducing and fixing a given $r_b$ (300 mm, 500 mm, 1200 mm and ∞) at room temperature. In Figure 3b, we show the normalized Kerr signal, $M/M_{185°C}$ (T), which is roughly proportional to the longitudinal magnetic moment of the film, collected at various $r_b$. We note that the FM to AFM transition takes place for all films, regardless of the bending state. The residual FM fractions in the AFM region are identical, and bending does not appreciably change the Néel temperature of the films. We also include in Figure 3b the magnetization versus temperature data of an unbent ($r_b = \infty$) film, measured by VSM.

In Figure 3c,d and e, we show the MOKE hysteresis loops collected while the sample is bent at different radii. Data have been recorded at some selected temperatures: $T_1 = 37°C$, $T_2 = 93°C$ and $T_3 = 185°C$, which are representative of the AFM, AFM-to-FM transition region and FM states, respectively. It can be observed that the loops recorded at $T_2 = 93°C$ and $T_3 = 185°C$ (Figure 3d and 3e) are characteristic of a ferromagnetic material. In contrast, the MOKE measurements recorded at 37°C (Figure 3c) does not show any hysteretic behavior, reflecting that FeRh is mainly AFM at this temperature and the contribution of residual FM nuclei are within the noise level.

It can be appreciated that the loops recorded deep in the FM state ($T_3$) (Figure 3e) exhibit a MOKE signal, insensitive to bending, slightly larger than that observed in the loops recorded at the verge of the FM-to-AFM transition ($T_2$). At the intermediate temperature $T_2$ (Figure 3d), the ferromagnetic loop is clearly visible and only a marginal reduction of magnetization with bending ascribed to the fact that the bending-induced strain is relatively small ($< 0.02\%$)[57-60] (see Supporting Information Appendix S1). In short, bending does no produce appreciable changes of film magnetization, when the films are deep in the FM (high temperature, $T_3$) or AFM phase (low temperature, $T_1$).

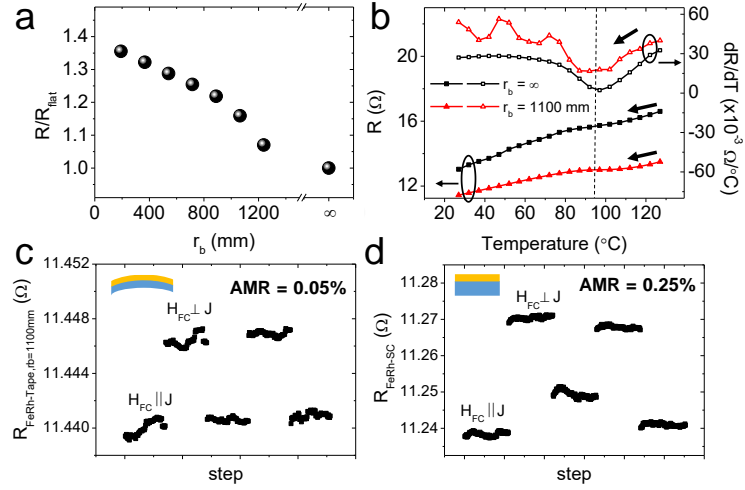

**Figure 4.** (a) Dependence of the electrical resistance of the FeRh-Tape film on the bending (convex $r_b$) recorded at room temperature. The relative variation $R(r_b)/R(\infty)$ is depicted. (b) Left-axis: Resistance dependence on temperature for the unbent state ($r_b = \infty$) and under bending ($r_b = 1100$ mm). Right-axis: Derivative of the resistance versus temperature. Resistance measured at room-temperature (26.85°C) and zero magnetic field after field-cooling ($H_{FC} = 90$ kOe) process of: (c) bent FeRh-tape and (d) FeRh-SC, with the current (J) parallel (for the low resistance state) or perpendicular (for the high resistance state) to cooling field ($H_{FC}$) as indicated.

The effect of bending is visible in the film resistance as illustrated in **Figure 4**a. The room-temperature resistance $R(r_b)$ was recorded while bending the film and the relative variation $R(r_b)/R(\infty)$ is plotted versus $r_b$ in Figure 4a. It can be observed that the resistance increases up to 40 % when bending. We have also analyzed the temperature dependence of the resistance $R(T)$ when FeRh-Tape samples are thermally cycled across the FM-AFM phase transition in the bent ($r_b = 1100$ mm) and unbent ($r_b = \infty$) states. Results are shown in Figure 4b. We first note that these data confirm that bending increases the film resistance, as anticipated in Figure 4a. Next, a detailed inspection of Figure 4b reveals a change of slope of $R(T)$ which signals the Néel temperature $T_N$, is well visible in both sets of data.[48] This observation is more apparent in Figure 4b (right axis), where we plot the derivate of the film resistance [dR/dT(T)] versus temperature. Overall, data show that film bending does not modify $T_N$ but unbalances the AFM/FM fraction thus changing the film resistance. The smaller difference of resistance between FM and AFM phases in FeRh-tape compared to FeRh films grown on single crystal substrates[48] is related to the presence of a larger residual ferromagnetic fraction at room-temperature (within the antiferromagnetic region) in the FeRh tapes.

Finally, we address the capability of the bendable FeRh film developed here to store and retrieve magnetic information, by using the protocol developed by Marti et al.[28, 31] In short, data is written by selecting the magnetization direction of FeRh in its high-temperature ferromagnetic state by a suitable magnetic field, and cooling it under field to room-temperature. After zeroing the magnetic field, data (Néel vector direction) is stored in the antiferromagnetic state. Data is retrieved at room-temperature by measuring the electric resistance of the film in two perpendicular configurations (parallel and perpendicular to the cooling filed direction). In Figure 4c, the resistance (R) of the FeRh-Tape (bent to $r_b$ = 1100 mm) recorded with the current direction (J) parallel and perpendicular to the zeroed cooling field ($H_{FC}$) is shown. An obvious resistance contrast [between R($H_{FC}\perp J$) and R($H_{FC}\|J$)] is visible illustrating the data storing and retrieval capabilities of bendable FeRh tapes. Resistance contrast (AMR) of about 0.05%, is found to be comparable although smaller than in FeRh alloys grown on single crystalline substrates (Figure 4d). The difference may be related to the presence of a larger residual (low temperature) ferromagnetic phase in the FeRh tapes compared to films on single crystalline substrates. Growth optimization procedures may allow to recover larger AMR values, eventually mimicking FeRh-SC. Device engineering can be also used for optimized reading, for instance, by incorporating a tunnel sensor, the magnetoresistance can be amplified up to 15%.[61]

## 3. Conclusions

In summary, we have taken advantage of the availability of MgO-buffered flexible C267 HASTELLOY® long tapes developed for high $T_c$ superconductors technologies, to demonstrate that the $\alpha$-FeRh alloy can be successfully grown on technologically affordable and flexible substrates and used to store and retrieve magnetic information in its antiferromagnetic phase. The FeRh tapes display the AFM/FM transition close to room temperature and thus all potential benefits already claimed for crystalline FeRh grown on single-crystalline substrates, can now be translated and expanded to meter-long and flexible substrates. Our results indicate that upon bending, the room-temperature AFM and the high-temperature FM phases remain stable. Therefore, applications exploiting the AFM/FM transition of FeRh, are now possible on flexible FeRh tapes. Moreover, we have demonstrated that bendable FeRh tapes can be used to store magnetic information, which is recalled by reading its electrical resistance. New opportunities and applications can be foreseen for these flexible memory elements. For instance, cloaked magnetic information on wearable or flexible supports.

## 4. Experimental section

**Sample growth.** FeRh (57 nm) films were grown on a textured MgO buffer ($\approx 2$ μm thick) prepared by ISD (Inclined Substrate Deposition) technique on a Nickel-Molybdenum-Chromium alloy (C267 HASTELLOY®) flexible tape ($\approx 100$ μm-thick, Theva Dünnschichttechnik GmbH). The FeRh layer was protected *in-situ* by a 20nm-Pt coating as sketched in Figure 1a. The MgO-C267 tape with 25 mm lenght and 12 mm width was cut into 10 mm × 12 mm pieces and fixed into the substrate heater of the sputtering chamber by using Ag paste. The tape substrate was heated up to the deposition temperature $T_D = 300°$ in a pressure of 0.01 mbar of Ar. The FeRh film was grown by DC magnetron sputtering using 20 W power at a rate of 0.8 Å/s from a stoichiometric target. Afterwards the film was annealed at $T_A = 700°C$ in a pressure of 0.1 mbar of Ar for $t_A = 1$ h and subsequently cooled to room temperature at a 100ºC/min rate. Finally, at room temperature 20 nm of Pt was deposited by sputtering in 0.005 mbar of Ar atmosphere using 20 W power. The control sample grown on MgO single crystal was grown using the same conditions. Sample shown in Figure 4(d) is 100 nm and without Pt capping to optimize AMR magnitude. Other samples were also grown on flexible tape using slightly different conditions, as summarized in Supporting Information Table S1.

**X-ray diffraction.** The crystal structure was analyzed by X-ray diffraction (XRD) using Cu $K_\alpha$ radiation. 2D diffraction $2\theta-\psi$ frames were recorded using Bruker D8 Advance diffractometer with 2D detector. The integrated along Ψ scans (Figure 1d) are integrated for Ψ between -15º and +15º of Figure 2h for FeRh-SC and for Ψ between the enclosed dashed region of Figure 2g for FeRh-Tape.

**Morphological characterization.** SE and BS SEM images were recorded using a Fei Quanta 200EF microscope. FeRh-SC sample was cleaved and FeRh-Tape sample was cut and samples were imaged with a tilting angle of ≈80 ° with respect the normal film direction for near grazing incidence images. Composed (SE+BS) images were the result of the sum of SE and BE images to enhance MgO and FeRh contrast.

**Magnetic characterization.** Magnetization versus temperature characteristics were measured by Vibrating Sample Magnetometer (VSM) platform from Quantum Design at H = 500 Oe, along the in-plane direction.

**Magneto-optic Kerr effect characterization.** The evolution of the local magnetic properties were investigated using a Durham Magneto Optics Ltd polar magneto-optic Kerr effect (MOKE) apparatus with a laser spot focused down to around 3 μm. Temperature was extracted from in-situ Nickel-Molybdenum-Chromium alloy resistance measurement using resistance versus temperature calibration data collected in a Physical Properties Measurement System (PPMS). The MOKE experiments were performed applying an in-plane magnetic field ($H_{app}$ < 500 Oe). MOKE data for bent state correspond to the convex bending side of the sample. The amplitude of the Kerr rotation $\Theta_K$ (i.e., MOKE signal) is roughly proportional to the longitudinal magnetic moment of the film (m) and ultimately to its magnetization (M): $\Theta_K(H_{app}) \propto m(H_{app}) \propto M(H_{app})$. For each $r_b$, the MOKE data were normalized the maximum Kerr rotation amplitude which is achieved at 185 °C where the alloy is fully ferromagnetic (that is, $\Theta_K(H_{app},T)/\Theta_K(H_{app,185\ °C}) = M(H_{app},T)/M_{185°C}$].

**XMCD-PEEM characterization.** XMCD–PEEM experiments were performed at the CIRCE beamline of the ALBA Synchrotron[62] using circularly polarized x-ray with an energy resolution of $E/\Delta E \approx 5000$. Before imaging the Pt capping layer was removed to obtain better magnetic contrast. All images were recorded at the Fe-$_{L3}$ edge at ≈ 707 eV. The field of view was 10 μm.

**Electric transport characterization and anisotropic magnetoresistance.** The resistance versus temperature measurements were performed in two-probe configuration with a constant current of 100 mA while cooling the sample at 2 °C/min in a Physical Properties Measurement

System (PPMS). The protocol followed in the AMR measurements of Figures 4(c,d) is: 1) field cool ($H_{FC}$ = 90kOe) down to room-temperature (26.85°C) along perpendicular and parallel to the measuring current directions, 2) set the magnetic field to zero and 3) measure the resistance. In Figures 4(c,d), each data point for each state correspond to consecutively recorded measurements with a delay time of ≈1s, at the same temperature and after the same field cooling procedure. In Figures 4(c,d), the different resistance values were obtained after applying the mentioned field cooling protocol. The resistance contrast corresponds to AMR = [R($H_{FC}\perp$J) - R($H_{FC}\|$J)]/ R($H_{FC}\|$J).

## ASSOCIATED CONTENT

The Supporting Information is available free of charge at

Complementary structural, morphological and magnetic characterization (PDF).

## AUTHOR INFORMATION

**Corresponding Author**

*ifina@icmab.es, *fontcuberta@icmab.cat

## ACKNOWLEDGMENT

Financial support from the Spanish Ministry of Economy and Competitiveness, through the "Severo Ochoa" Programme for Centres of Excellence in R&D (SEV-2015-0496) and the RTI2018-095303-B-C53, MAT2017-85232-R, MAT2014-56063-C2-1-R, and MAT2015-73839-JIN projects, the Generalitat de Catalunya (2017 SGR 1377) and the European Union's Horizon 2020 research and innovation programme under the Marie Skłodowska-Curie grant agreement nº 665919 is acknowledged. I.F. acknowledges RyC contract RYC-2017-22531. M. Vilardell is acknowledged for advice in XRD characterization. Patxi López-Barberá is acknowledged for assistance in MOKE experiments. J. Oró for technical assistance in SEM characterization. We thank J. Nogués for the use of the MOKE equipment. V. García-Juez from Real Casa de la Moneda – Fábrica Nacional de Moneda y Timbre is acknowledged for scientific advice. Theva Dünnschichttechnik GmbH is acknowledged for providing the flexible substrates.

# Supporting Information: Flexible antiferromagnetic FeRh tapes as memory elements

Ignasi Fina,[†,*] Nico Dix,[†] Enric Menéndez,[‡] Anna Crespi,[†] Michael Foerster,[§] Lucia Aballe,[§] Florencio Sánchez,[†] Josep Fontcuberta[†,*]

[†]Institut de Ciència de Materials de Barcelona (ICMAB-CSIC), Campus UAB, E-08193 Bellaterra, Catalonia, Spain.

[‡]Departament de Física, Universitat Autònoma de Barcelona, E-08193 Bellaterra, Catalonia,, Spain.

[§]ALBA Synchrotron Light Facility, Carrer de la Llum 2-26, Cerdanyola del Vallès, Barcelona 08290, Spain.

*ifina@icmab.es, *fontcuberta@icmab.cat

**Table S1.** Summary of the characterized samples.

| Sample | $T_D$ (°C) | $T_A$ (°C) | $t_A$ (hours) |
|---|---|---|---|
| FeRh-tape | 300 | 700 | 1 |
| Different $T_D$ | 350 | 700 | 1 |
| Different $T_A$ | 300 | 750 | 1 |
| Different $t_A$ | 300 | 700 | 4 |

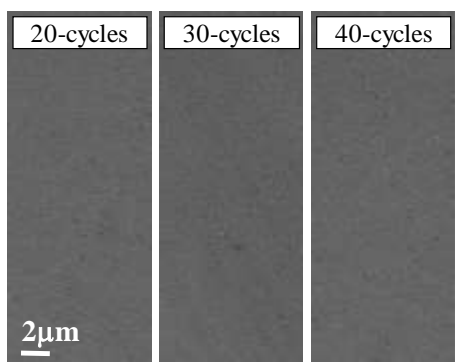

**Figure S1.** SE images of representative FeRh-Tape region after indicated number of bending cycles up to $r_b$ = 200 mm. No differences are observed.

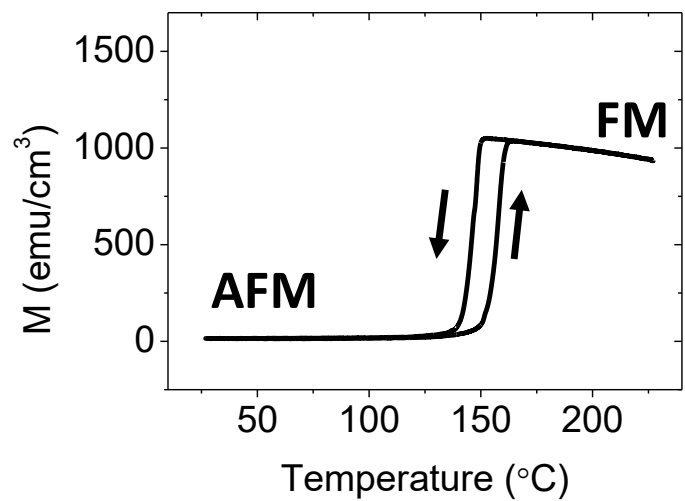

**Figure S2.** M(T) cycle collected at 500 Oe from 300 to 500 to 300 K following the arrows directions of a FeRh film grown on top of a single MgO(001) single-crystal. Sharp AFM-FM transition is observed, and it is used as reference for comparison with the FeRh-tape.

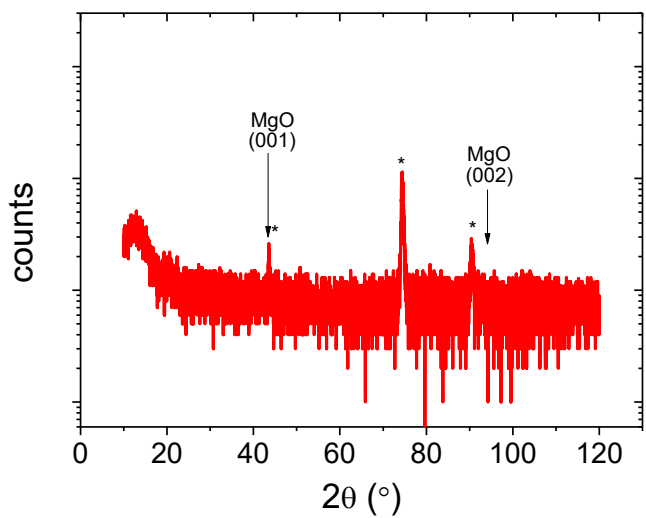

**Figure S3.** θ–2θ scan performed with point detector (Normal to the surface (ψ = 0) in the FeRh-Tape sample. HASTELLOY reflections are labelled with an (*).

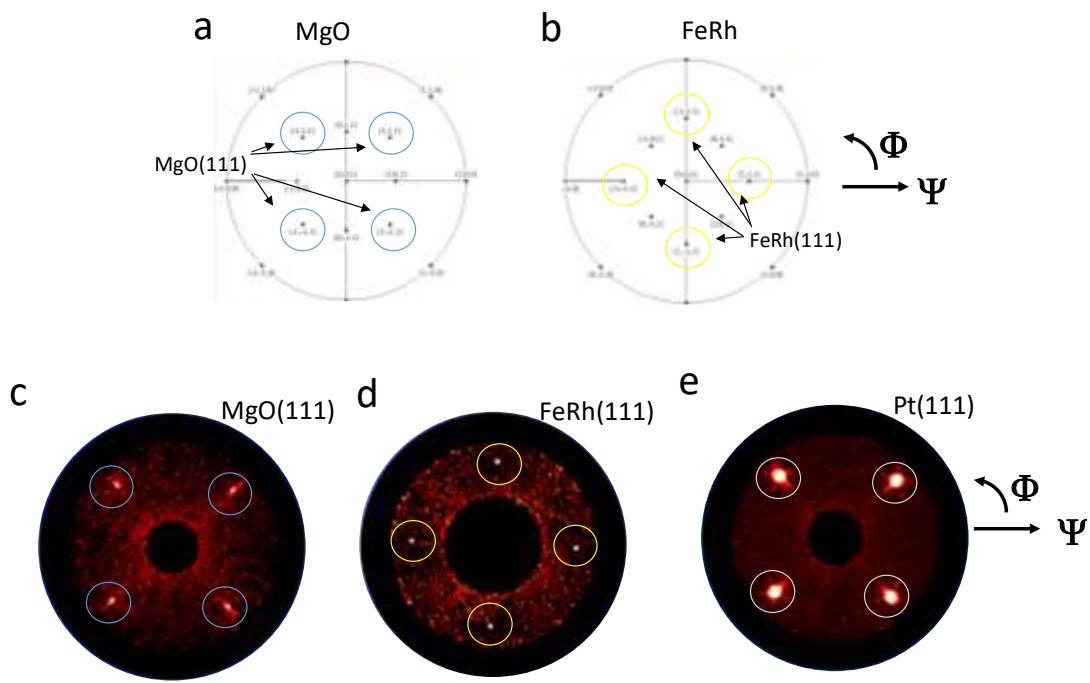

**Figure S4.** Simulated (using Carine software) pole figures of (a) MgO and Pt and (b) FeRh for different reflections considering [100]Pt(001)//[110]FeRh(001)//[100]MgO(001) epitaxial relation. Experimental pole figures of (c) MgO(111), (d) FeRh(111), and (e) Pt(111). The sharp and intense peaks are a result of the high crystalline quality of substrate and FeRh layer.

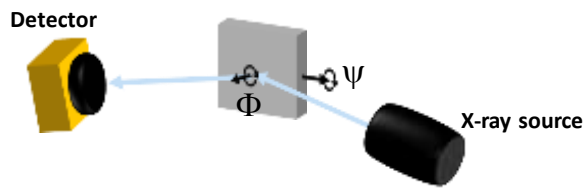

**Figure S5.** Sketch of the XRD set-up using a 2D detector. The angles $\Psi$ and $\Phi$ are shown. It can be observed that $\Psi$ tilting and $\Phi$ rotation can allow the alignment of the MgO and FeRh textured axes of the tape sample with the source detector configuration.

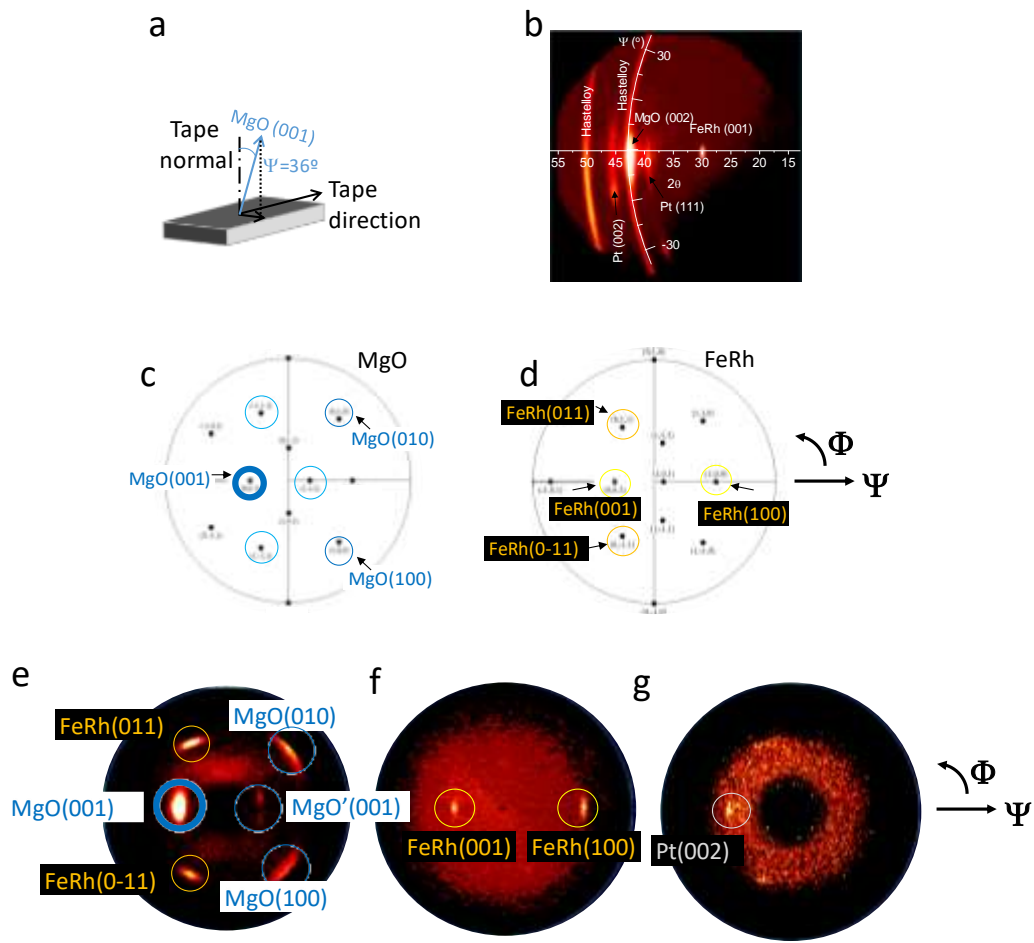

**Figure S6.** (a) Sketch of the MgO main tilted direction respect to the tape direction. (b) 2θ–Ψ frame (with sample goniometer at Ψ = 36º, φ=270º). Simulated (using Carine software) pole figures of (c) MgO and Pt and (d) FeRh for different reflections considering the [100]Pt(001)//[110]FeRh(001)//[100]MgO(001) epitaxial relationship and the texture shown in panel (a). Experimental pole figures (with sample goniometer at Ψ = 0º, φ=270º) of (e) MgO(002) and FeRh(110) and MgO'(002), (f) FeRh(001), and (g) Pt (002).

In (a) the sketch of the MgO tilting angle respect to the tape direction is shown. In (b) 2θ–Ψ frame centered in Ψ (with sample goniometer at Ψ = 36º, φ=270º) is shown. Here the MgO(002) and FeRh(002) reflections are clearly visible. There are Pt(002) and the less intense Pt(111) reflections elongated along Ψ, indicating important mosaicity. The hastelloy polycrystalline lines are also observed. In (e), the main MgO reflection is 001 (marked in thick blue circle) at around Ψ=36º of the surface normal. The pole figure also shows FeRh(110) and MgO'(001) reflections. In (f), the pole figure of FeRh shows (001) and (100) reflections. In (g), the main Pt(001) spot indicates epitaxy but with important degree of polycrystallinity, as expected since it is grown at room temperature. Finally, in panels (c,d), the shown simulations are in agreement with the found peak positions shown by the experimental data in panels (e,f,g) and confirm the [100]Pt(001)//[110]FeRh(001)// [100]MgO(001) epitaxial relationship as for the FeRh-SC sample.

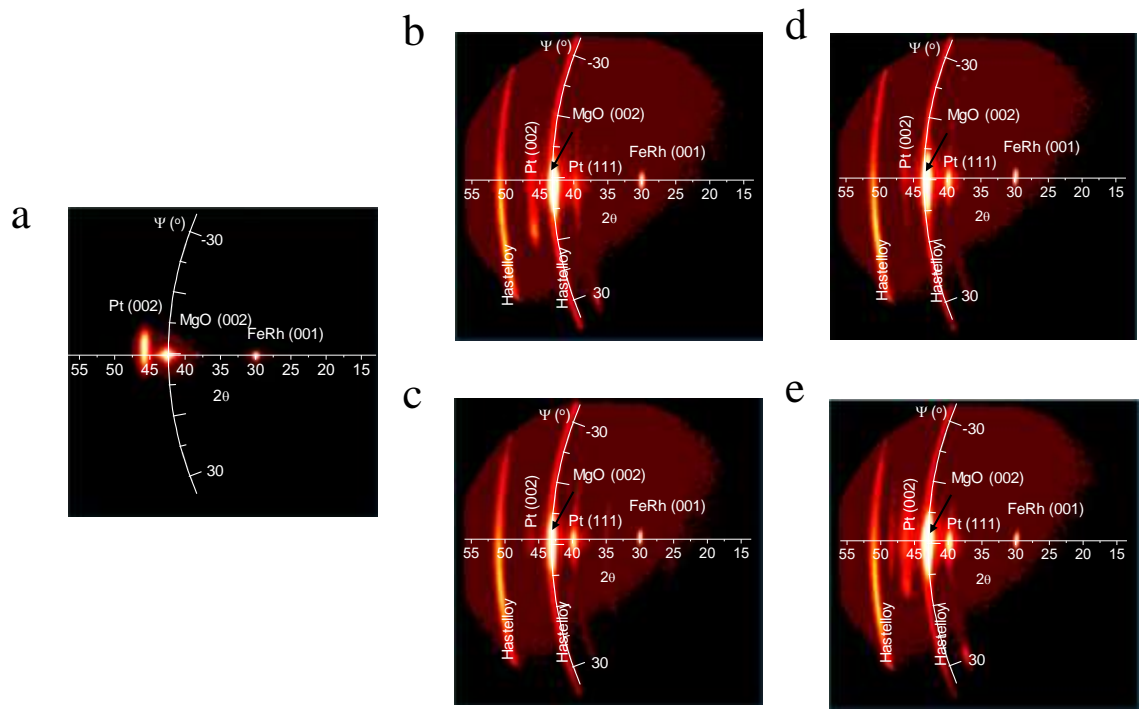

**Figure S7.** (a) 2D X-ray images along θ-2θ and Ψ directions centered along Ψ = 0° of FeRh-SC. 2D X-ray images along θ-2θ and Ψ directions centered along Ψ = 36° of (a) FeRh-SC, (b) FeRh-tape, (c) FeRh-tape annealed at higher (750°C) temperature, (d) FeRh-tape annealed for longer (4 hours) time, and (e) FeRh-tape growth at higher (350°C) temperature. No remarkable differences are observed.

Note that the Pt(002) reflection, occurring at Ψ ≈ -16º is beyond the 2 theta range of Figure 2g and thus not visible. Instead, the corresponding Pt(111) appears at Ψ ≈ 36º, as observed in Figure 2g. The Pt(002) reflection becomes apparent in the Ψ ≈ -16º centered images shown here.

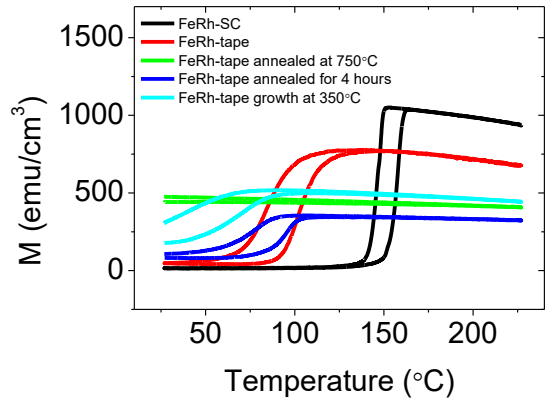

**Figure S8.** M(T) for FeRh-tape, FeRh-tape annealed at higher (750°C) temperature, FeRh-tape annealed for longer (4 hours) time, and FeRh-tape growth at higher (350°C) temperature. It can be observed that all the variations of the growth conditions of the FeRh-tape sample referred in the main text do not result in better functional properties, and both the magnetization at high temperature and the transition temperature are always smaller.

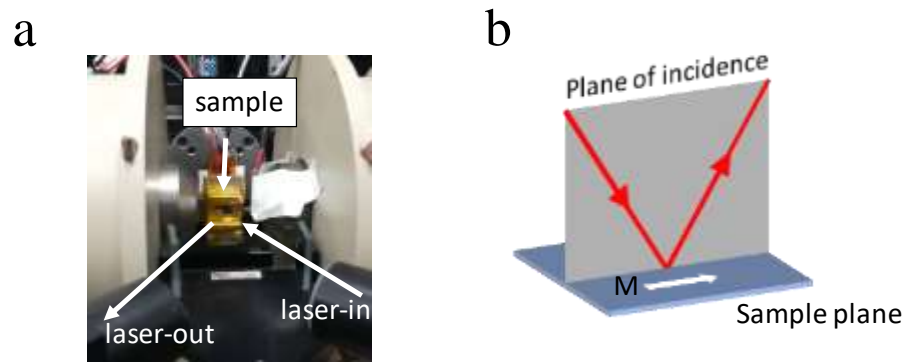

**Figure S9.** (a) Experimental set-up used for MOKE characterization. (b) Used longitudinal geometry.

*Appendix S1.* Using classical elastic theories, the in-plane strain induced on the film of thickness $t$ grown a substrate, by a bending of radius $R$ can be estimated by using

$$\varepsilon = \left(\frac{y+t}{R}\right)$$

Where $y$ is the distance from the neutral line of the whole FeRh/MgO/Hastelloy heterostructures. As the thickness of $t$(FeRh) ($\approx$ 50 nm) is much smaller than that of the MgO/Hastelloy substrate ($\approx$102 μm), then y $\approx$ 102 μm/2 = 51 μm) and thus the substrate imposes a roughly homogenous compressive stress on the FeRh of about $\varepsilon$ = 0.017% for R = 300 mm.